\documentclass[11pt,epsf]{article}
\usepackage{amsmath}
\usepackage{amsfonts}
\usepackage{amssymb}
\usepackage{graphicx}

\newcommand {\exe} {\stackrel{\cdot} {=}}
\newcommand {\dfn} {\stackrel{\Delta} {=}}

\newcommand {\reals} {{\rm I\!R}}

\newcommand {\bx} {\mbox{\boldmath $x$}}
\newcommand {\by} {\mbox{\boldmath $y$}}

\newcommand {\bE} {\mbox{\boldmath $E$}}

\newcommand {\bX} {\mbox{\boldmath $X$}}

\newcommand{\calA}{{\cal A}}
\newcommand{\calB}{{\cal B}}
\newcommand{\calC}{{\cal C}}
\newcommand{\calD}{{\cal D}}
\newcommand{\calE}{{\cal E}}

\newcommand{\calW}{{\cal W}}
\newcommand{\calX}{{\cal X}}
\newcommand{\calY}{{\cal Y}}

\begin{document}
\thispagestyle{empty}
\title{Relations Between Random Coding Exponents
and the Statistical Physics of Random Codes
\thanks{This research is partially supported by the
Israel Science Foundation (ISF), grant no.\ 223/05.
Part of this work was carried out during a visit at
HP Laboratories, Palo Alto, CA, U.S.A.\ in the Summer
of 2007.}}

\author{Neri Merhav
}
\maketitle

\begin{center}
Department of Electrical Engineering \\
Technion - Israel Institute of Technology \\
Haifa 32000, ISRAEL \\
\end{center}
\vspace{1.5\baselineskip}
\setlength{\baselineskip}{1.5\baselineskip}

\begin{abstract}
The partition function pertaining to finite--temperature decoding
of a (typical) randomly chosen code is known to have three types of behavior,
corresponding to three phases in the plane of rate vs.\ temperature:
the {\it ferromagnetic phase}, corresponding to correct decoding, the
{\it paramagnetic phase}, of complete disorder, which is 
dominated by exponentially many
incorrect codewords, and the {\it glassy phase} (or the condensed phase), where
the system is frozen at minimum energy and dominated by subexponentially
many incorrect codewords.
We show that the statistical physics associated with the two latter phases
are intimately related to random coding exponents. In particular, 
the exponent associated with the probability of correct decoding
at rates above capacity is directly related to the free energy in the
glassy phase, and the exponent associated with probability of error
(the error exponent) at rates below capacity, 
is strongly related to the free energy in the
paramagnetic phase. In fact, we derive alternative 
expressions of these exponents in terms
of the corresponding free energies, and make an
attempt to obtain some insights from these expressions.
Finally, as a side result, we also compare the phase
diagram associated with a simple finite--temperature 
universal decoder, for discrete
memoryless channels, to that of the finite--temperature decoder that 
is aware of the channel statistics.

\vspace{0.25cm}

\end{abstract}
\clearpage

\section{Introduction}

In the last few decades it has become apparent that many problems in
Information Theory, and the channel coding problem in particular,
can be mapped onto (and interpreted as) analogous problems in the
area of statistical physics of disordered systems (such as spin glass models).
Such analogies are useful because physical insights, as well as 
statistical mechanical tools and analysis techniques (like the replica method),
can be harnessed in order to advance the knowledge and the 
understanding with regard to 
the information--theoretic problem under discussion.
A very small, and by no means exhaustive, 
sample of works along this line includes references [1]--[29].

In this paper, we shall also adopt the 
statistical mechanical viewpoint on channel coding.
We focus on the classical random code ensemble (RCE) for
communicating over a discrete memoryless channel (DMC), 
in the same setting as described in
\cite[Chap.\ 6]{MM06} and \cite{MU07}, which in a nutshell, is as follows: 
Consider a DMC, $P(\by|\bx)=\prod_{i=1}^n p(y_i|x_i)$, fed by an input
$n$--vector that belongs to
a codebook $\calC=\{\bx_1,\bx_2,\ldots,\bx_M\}$, $M=e^{nR}$,
with uniform priors, where $R$ is the coding rate in nats per channel use.
The induced posterior, for $\bx\in\calC$, 
is then:
\begin{eqnarray}
P(\bx|\by)&=&\frac{P(\by|\bx)}{\sum_{\bx'\in\calC}P(\by|\bx')}\nonumber\\
&=&\frac{e^{-\ln[1/P(\by|\bx)]}}
{\sum_{\bx'\in\calC}e^{-\ln[1/P(\by|\bx')]}}.
\end{eqnarray}
Here, the second line is written in a form that resembles the
Boltzmann distribution of statistical physics, according to
which the probability of a certain `state' (or `configuration') 
of the system, designated by $\bx$,
is given by
\begin{equation}
\label{boltzmann}
P(\bx)= \frac{e^{-\beta \calE(\bx)}}{Z(\beta)}
\end{equation}
where $\beta=1/(kT)$ is the inverse temperature, $k$ is Boltzmann's
constant,\footnote{Here we will adopt the convention, customarily used
in many papers and books, 
of redefining `temperature' according to $T\leftarrow kT$, that is,
in units of energy, and then
$\beta\dfn 1/T$.}
$T$ is temperature, $\calE(\bx)$ is the energy associated with $\bx$,
and $Z(\beta)=\sum_{\bx}
e^{-\beta \calE(\bx)}$ is the {\it partition function}.
In our case, of course, $\beta=1$ and 
the energy function (which depends on the given $\by$) is
$\calE(\bx)=\ln[1/P(\by|\bx)]$. But this analogy with the
Boltzmann distribution (\ref{boltzmann}) naturally suggests (cf.\ e.g., 
\cite{MM06}) to
consider, more generally, the posterior distribution parametrized
by $\beta$, that is
\begin{eqnarray}
\label{pbeta}
P_\beta(\bx|\by)&=&\frac{P^\beta(\by|\bx)}{\sum_{\bx'\in\calC}
P^\beta(\by|\bx')}\nonumber\\
&=&\frac{e^{-\beta\ln[1/P(\by|\bx)]}}{\sum_{\bx'\in\calC}
e^{-\beta\ln[1/P(\by|\bx')]}}\nonumber\\
&\dfn&\frac{e^{-\beta\ln[1/P(\by|\bx)]}}{Z(\beta|\by)}.
\end{eqnarray}
There are a few motivations for introducing the temperature parameter
in (\ref{pbeta}). First, it allows a degree of freedom
in case there is some uncertainty regarding the channel 
noise level (small $\beta$ corresponds to high noise level). Second, it is
inspired by the ideas behind simulated annealing techniques: by
sampling from $P_\beta$ while gradually increasing $\beta$
(cooling the system), the minima of the energy function 
(ground states) can be found.
Third, by applying symbolwise MAP decoding, i.e., decoding the $\ell$--th
symbol of $\bx$ as 
$\mbox{arg}\max_a P_\beta(x_\ell=a|\by)$, where
$$P_\beta(x_\ell=a|\by)=
\sum_{\bx\in\calC:~x_\ell=a}P_\beta(\bx|\by),$$ 
we obtain
a family of 
{\it finite--temperature decoders} 
(originally proposed by Ruj\'an \cite{Rujan93}; see also \cite{Sourlas94},
\cite[Section 6.3.3]{MM06},\cite{Iba99},\cite{FLMR02})
parametrized by $\beta$, where $\beta=1$ corresponds
to minimum symbol error probability (with respect to the true channel)
and $\beta\to\infty$ corresponds to minimum block error probability.
Finally, and this is 
the motivation that drives the research reported in this paper:
the corresponding partition function, $Z(\beta|\by)$, namely,
the sum of (conditional) probabilities raised to some power $\beta$, is
an expression frequently encountered in 
R\'enyi information measures as well as
in the analysis of random 
coding exponents using Gallager's techniques.
Since the partition function plays a key role in statistical mechanics,
as many physical quantities can be derived from it,
then it is natural to ask if it can also be used to 
gain some insights regarding the behavior of random codes at various
temperatures and coding rates. 
The main contribution of this paper
is in exploring this direction.

To sharpen the last point a little further,
it is noted that when one considers the random coding regime,
as we do in this paper, then even if $\by$ is given, the energy levels
pertaining to the Boltzmann distribution (\ref{pbeta}) are themselves 
random variables since they depend on the randomly chosen codevectors.
As explained in \cite{MM06}, this then falls under the umbrella of the
so called {\it random energy model} (REM) in statistical physics, 
which was invented by Derrida
\cite{Derrida81} with the motivation 
to capture disorder in spin glass systems.
The interesting fact about the REM is that it is typically subjected to
{\it phase transitions}, and then so is the 
model (\ref{pbeta}) for random codes.

More specifically, as described in \cite[Chap.\ 6]{MM06},
\cite{Montanari01}, and as will be briefly reviewed in the next section,
the partition function pertaining to finite--temperature decoding
of a (typical) randomly chosen code is known to have three types of behavior,
corresponding to three phases in the plane of rate vs.\ temperature:
the {\it ferromagnetic phase}, corresponding to correct decoding, the
{\it paramagnetic phase}, of complete disorder, which is 
dominated by exponentially many
incorrect codewords, and the {\it glassy phase} (or the condensed phase), where
the system is frozen at minimum energy and dominated by subexponentially
many incorrect codewords.
We show that the statistical physics associated with the two latter phases
are intimately related to random coding exponents. In particular, 
the exponent associated with the probability of correct decoding
at rates above capacity is directly related to the free energy in the
glassy phase, and the exponent associated with probability of error
(the error exponent) at rates below capacity, 
is strongly related to the free energy in the
paramagnetic phase. In fact, we derive alternative 
expressions of these exponents in terms
of the corresponding free energies, and make an
attempt to obtain some insights from these expressions.

An additional interesting byproduct of the statistical mechanical point of
view that we adopt in this work, is that it suggests a more refined
analysis technique, as an alternative to the customary
use of Jensen's inequality,
for which it is clear
that the resulting expressions are exponentially tight, and not just bounds.
Another way to look at this is to observe
that the analysis technique, inspired
by statistical mechanical point of view, provides us with insights
with regard to the conditions under which Jensen's inequality provides
a tight bound in this context. We believe that
this technique may be useful in other applications as well.
We shall elaborate more on this in the sequel.

As a side result, we also compare the phase
diagram associated with a certain 
universal decoder (namely, the minimum conditional entropy universal
decoder) for discrete
memoryless channels, to that of the finite--temperature decoder that 
is aware of the channel statistics, and show that in spite of the
fact that this universal decoder is asymptotically optimum, in the
sense of attaining optimum random coding 
error exponents \cite{Ziv85}, its phase
diagram is substantially different.

The outline of the remaining part of this paper is as follows.
In Section 3, we provide some background, which mostly follows
the presentation in \cite{MM06} (with a few missing details filled in),
but will be useful here to keep this
paper self contained. Section 3 also includes a subsection with
the phase diagram for universal decoding, as described in the previous
paragraph. In Section 4, we derive the alternative formula for
the exponent of correct decoding above capacity, and in Section 5,
we do the same regarding the random coding exponent at rates below
capacity.

\section{Notation Conventions, Background and Preliminaries}

\subsection{Notation Conventions}

Throughout this paper, scalar random 
variables (RV's) will be denoted by capital
letters, like $X$ and $Y$, their sample values will be denoted by
the respective lower case letters, and their alphabets will be denoted
by the respective calligraphic letters.
A similar convention will apply to
random vectors and their sample values,
which will be denoted with the same symbols in the boldface font.
Thus, for example, $\bX$ will denote a random $n$-vector $(X_1,\ldots,X_n)$,
and $\bx=(x_1,...,x_n)$ is a specific vector value in $\calX^n$,
the $n$-th Cartesian power of $\calX$. 

Sources and channels will be denoted generically by the letters $P$ and $Q$.
Specific letter probabilities corresponding to a source $Q$ will be
denoted by the corresponding lower case letters, e.g., $q(x)$ is the
probability of a letter $x\in\calX$. A similar convention will be applied
to the channel $P$ and the 
corresponding transition probabilities, $p(y|x)$,
$x\in\calX$, $y\in\calY$. The expectation operator will be
denoted by $\bE\{\cdot\}$.

The empirical distribution pertaining
to a vector $\bx\in\calX^n$ will be denoted by $\hat{P}_{\bx}$.
In other words, $\hat{P}_{\bx}=\{\hat{p}_{\bx}(a),~a\in\calX\}$, where
${p}_{\bx}(a)=n_{\bx}(a)/n$, $n_{\bx}(a)$ being the number of occurrences
of the letter $a$ in $\bx$. Similar conventions will apply to
empirical joint distributions of pairs of letters, $(a,b)\in\calX\times\calY$,
extracted from the corresponding pairs of vectors $(\bx,\by)$, that is,
the joint empirical distribution $\hat{P}_{\bx\by}$ is the vector
of relative frequencies of joint occurrences of $x_i=a$ and $y_i=b$,
$i=1,\ldots,n$. Similarly, $\hat{p}_{\bx|\by}(a|b)=\hat{p}_{\bx\by}(a,b)/
\hat{p}_{\by}(b)$ will denote the empirical conditional probability of
$X=a$ given $Y=b$ (with convention that $0/0=0$), 
and $\hat{P}_{\bx|\by}$ will denote
$\{\hat{p}_{\bx|\by}(a|b),~a\in\calX,~b\in\calY\}$.
The expectation w.r.t.\ the empirical distribution of $(\bx,\by)$
will be denoted by $\hat{\bE}_{\bx\by}\{\cdot\}$, i.e., for a given
function $f:\calX\times\calY\to\reals$, we define
$\hat{\bE}_{\bx\by}\{f(X,Y)\}$ as $\sum_{(a,b)\in\calX\times\calY}
\hat{p}_{\bx\by}(a,b)f(a,b)$, where in this notation, $X$ and $Y$
are understood to be random variables jointly distributed according
to $\hat{P}_{\bx\by}$.

The cardinality of a finite set $\calA$ will be denoted by $|\calA|$.
For two positive sequences $\{a_n\}$ and $\{b_n\}$, the notation
$a_n\exe b_n$ means that $a_n$ and $b_n$ are asymptotically of the same
exponential order, that is, $\lim_{n\to\infty}\frac{1}{n}\ln\frac{a_n}{b_n}
=0$. Information theoretic quantities like entropies and mutual
informations will be denoted following the usual conventions
of the Information Theory literature. When we wish to make it clear
that such an information theoretic quantity is induced by a certain
probability distribution, say $Q$, we use this probability distribution as a
subscript, e.g., $I_Q(X;Y)$, $H_Q(X|Y)$, etc. When the underlying
probability distribution is an empirical distribution, we will subscript
it by the sequences(s) from which the empirical distribution is extracted, and
we will use hats, e.g., $\hat{I}_{\bx\by}(X;Y)$, $\hat{H}_{\bx\by}(X|Y)$.

\subsection{Background and Preliminaries}

Consider a DMC with a finite input alphabet $\calX$ and a finite output
alphabet $\calY$, which when fed by an input vector $\bx\in\calX^n$,
it generates an output vector $\by\in\calY^n$ distributed according to
$$P(\by|\bx)=\prod_{i=1}^n p(y_i|x_i),$$
where $\{p(y|x),~x\in\calX,~y\in\calY\}$ are given single--letter
transition probabilities. Let 
$$\calC=\{\bx_1,\bx_2,\ldots,\bx_M\}\subseteq
\calX^n$$ 
be a codebook of $M=e^{nR}$ codewords, where $R$ is the coding
rate (in nats per channel use). Next consider the posterior distribution
(\ref{pbeta}) and the corresponding partition function
\begin{equation}
Z(\beta|\by)=\sum_{\bx\in\calC}P^\beta(\by|\bx)
=\sum_{\bx\in\calC}e^{-\beta d(\bx,\by)},
\end{equation}
where $d(\bx,\by)=-\ln P(\by|\bx)=-\sum_{i=1}^n\ln P(y_i|x_i)$.
We shall think of $Z(\beta|\by)$ as the sum of two contributions,
the first is $Z_c(\beta|\by)=
e^{-\beta d(\bx_0,\by)}$, pertaining to the correct codeword $\bx_0$
(that was actually transmitted across the channel), and the second is
associated with the remaining (incorrect) codewords,
$$Z_e(\beta|\by)=\sum_{\bx\in\calC-\{\bx_0\}}e^{-\beta d(\bx,\by)}.$$
Let us focus on $Z_e(\beta|\by)$ first. As mentioned in the Introduction,
when the codebook $\calC$ is selected at random, this is 
a disordered system in the framework of the REM,
which exhibits phase transitions. 

To describe these phase transitions,
it is instructive to begin with the relatively simple special case of
the binary symmetric channel (BSC), as we do in Subsection 2.2.1, and
then extend the scope to general DMC's, as in Subsection 2.2.2.\footnote{
This extension to general DMC's in outlined in \cite[Chap.\ 6]{MM06}, but here
we provide some more details.} Finally, Subsection 2.2.3 (which is not
included in \cite{MM06}) is about 
a phase diagram pertaining to universal decoding (cf.\ second to 
the last paragraph of the Introduction). This subsection can be skipped
without loss of continuity.

\subsubsection{The Binary Symmetric Channel}

For the BSC with a crossover parameter $p$, we have
$P(\by|\bx)=p^{d_H(\bx,\by)}(1-p)^{n-d_H(\bx,\by)}$,
where $d_H(\bx,\by)$ is the Hamming distance between $\bx$ and $\by$. 
Defining $B=\ln\frac{1-p}{p}$, we then
have $P(\by|\bx)=(1-p)^n e^{-Bd_H(\bx,\by)}$, and so
\begin{eqnarray}
Z_e(\beta|\by)&=&\sum_{\bx\in\calC}P^\beta(\by|\bx)\nonumber\\
&=&(1-p)^{\beta n}\sum_{\bx\in\calC}e^{-\beta Bd_H(\bx,\by)}\nonumber\\
&=&(1-p)^{\beta n}\sum_{d=0}^nN_{\by}(d)e^{-\beta Bd},
\end{eqnarray}
where $N_{\by}(d)$ is the number of 
incorrect codewords at Hamming distance $d$ from $\by$.
As argued in \cite{MM06}, when the codewords are chosen independently
at random (say, by fair coin tossing),
$\{N_{\by}(d)\}$ concentrate very rapidly,\footnote{Note that $N_{\by}(d)=
\sum_{i=1}^{e^{nR}}1\{d_H(\bx_i,\by)=d\}$, i.e., it is the sum of
exponentially many i.i.d.\ (given $\by$) random variables, and so, its
large deviations behavior is
exponential in $e^{nR}$, which is double--exponential in $n$
(see also Appendix, Subsection A.2.).}
as $n\to\infty$, about
their expectations:
\begin{equation}
\bE\{N_{\by}(\delta n)\}\exe e^{n[R-\ln 2+h(\delta)]},~~~0\le\delta\le 1
\end{equation}
where $h(\delta)\dfn-\delta\ln\delta-(1-\delta)\ln(1-\delta)$.
Defining the normalized Gilbert--Varshamov (GV) distance,
$\delta_{GV}(R)$, as the solution, $\delta$, 
to the equation $h(\delta)=\ln 2-R$,
it is apparent that for $\delta < \delta_{GV}(R)$ 
and $\delta > 1-\delta_{GV}(R)$,
$\bE\{N_{\by}(\delta n)\}$ has a negative exponent, and thus typically, these
distances are not populated by codewords. Therefore, for a typical random code,
\begin{eqnarray}
Z_e(\beta|\by)&\exe&(1-p)^{\beta n}\cdot 
e^{n(R-\ln 2)}\int_{\delta_{GV})}^{1-\delta_{GV}(R)}
\mbox{d}\delta \cdot e^{nh(\delta)}\cdot e^{-\beta B\delta}\nonumber\\
&\exe&(1-p)^{\beta n}\cdot e^{n(R-\ln 2)}
\exp\left\{n\cdot\max_{\delta\in[\delta_{GV}(R),1-\delta_{GV}(R)]}\left[
h(\delta)-\beta B\delta\right]\right\}\nonumber\\
&\dfn& e^{-n\beta F_e(\beta)}
\end{eqnarray}
where $F_e(\beta)$ is the {\it free energy density}
associated with the incorrect codewords, which is
given by
\begin{equation}
F_e(\beta)=\left\{ \begin{array}{ll}
\delta_{GV}(R)\ln\frac{1}{p}+(1-\delta_{GV}(R))\ln\frac{1}{1-p} 
& p_\beta\le\delta_{GV}(R) \\
\frac{1}{\beta}[\ln 2-R-\ln(p^\beta+(1-p)^\beta)] & p_\beta > \delta_{GV}(R) 
\end{array}\right.
\end{equation}
where 
$$p_\beta=\frac{p^\beta}{p^\beta+(1-p)^\beta},$$
and where the distinction between the two different expressions is
due to the constraint $\delta\in[\delta_{GV}(R),1-\delta_{GV}(R)]$,
which becomes active (i.e., achieved with equality) when 
$p_\beta \le \delta_{GV}(R)$.
We observe then that
when $p$ and $R$ are held fixed, and $\beta$ varies, the above expression
exhibits a phase transition at temperature $T_c(R)=1/\beta_c(R)$
for which $p_\beta = \delta_{GV}(R)$, i.e.,
$$\beta_c(R)=\frac{\ln[(1-\delta_{GV}(R))/\delta_{GV}(R)]}{\ln[(1-p)/p]}.$$
For $\beta > \beta_c$ (low temperature), the free energy density
$F_e(\beta)=\delta_{GV}(R)\ln\frac{1}{p}
+(1-\delta_{GV}(R))\ln\frac{1}{1-p}$ is independent of
$\beta$ hence the entropy 
(which is related to the derivative of $F_e(\beta)$ w.r.t.\ $\beta$) vanishes,
and the system is frozen in the sense that the
thermodynamics are dominated by a subexponential number of configurations of
the minimum energy which is $n\delta_{GV}(R)$. This phase is referred to 
as {\it condensed phase} or {\it glassy phase},
and henceforth we denote
$$F_g\dfn\delta_{GV}(R)\ln\frac{1}{p}+(1-\delta_{GV}(R))\ln\frac{1}{1-p}.$$
For $\beta > \beta_c$, the thermodynamics are dominated 
by an exponential number of states
at distance $np_\beta$, which is 
larger than $n\delta_{GV}(R)$, and the entropy is strictly positive.
This is called the {\it paramagnetic phase} and henceforth we denote
$$F_p(\beta)\dfn\frac{1}{\beta}[\ln 2-R-\ln(p^\beta+(1-p)^\beta)].$$

When the contribution of $Z_c(\beta)=e^{-n\beta F_c}$ 
is taken into
account, and we consider the total partition 
function $Z(\beta)$, the situation changes:
Since $d_H(\bx_0,\by)$ is typically about 
the level of $np$, and thus the corresponding free energy density is 
$F_c=h(p)$,
we have yet another phase referred
to as the {\it ordered phase} or 
the {\it ferromagnetic phase}. This phase exists whenever
$Z(\beta)$ is dominated by $Z_c(\beta)$, i.e., $F_c=h(p) < F_e(\beta)$.
For $\beta > \beta_c$, this 
is the case whenever $p < \delta_{GV}(R)$, or equivalently,
$R < \ln 2-h(p)\dfn C$, where $C$ is the capacity of the BSC. 
For $\beta < \beta_c$ 
the boundary between the ferromagnetic phase
and the paramagnetic phase is given by 
the solution $\beta_0(R)=1/T_0(R)$ to the equation 
\begin{equation}
\label{ferropara}
\beta h(p)= \ln 2-R-\ln[p^\beta+(1-p)^\beta].
\end{equation}

To summarize, while there are only two phases 
(glassy and paramagnetic) pertaining to $Z_e(\beta)$,
there is a third, additional phase (ferromagnetic) associated with $Z_c(\beta)$.
In the ferromagnetic phase, the system is dominated by one state corresponding 
to the correct codeword. Thus, similarly as in the glassy phase, the entropy
of the ferromagnetic phase is zero. The boundaries between the three phases 
in the plane defined by $R$ and $T=1/\beta$, are
as follows (see Fig.\ 1): 
The ferro--glassy boundary is the straight line $R=C$,
the glassy--paramagnetic boundary is the curve $T=T_c(R)$, and the
and the ferro--paramagnetic boundary $T=T_0(R)$ 
is given by eq.\ (\ref{ferropara}).
The triple point where all boundaries intersect is the point 
$(R,T)=(C,1)$.

\begin{figure}[ht]
\hspace*{1cm}
\begin{picture}(0,0)%
\includegraphics{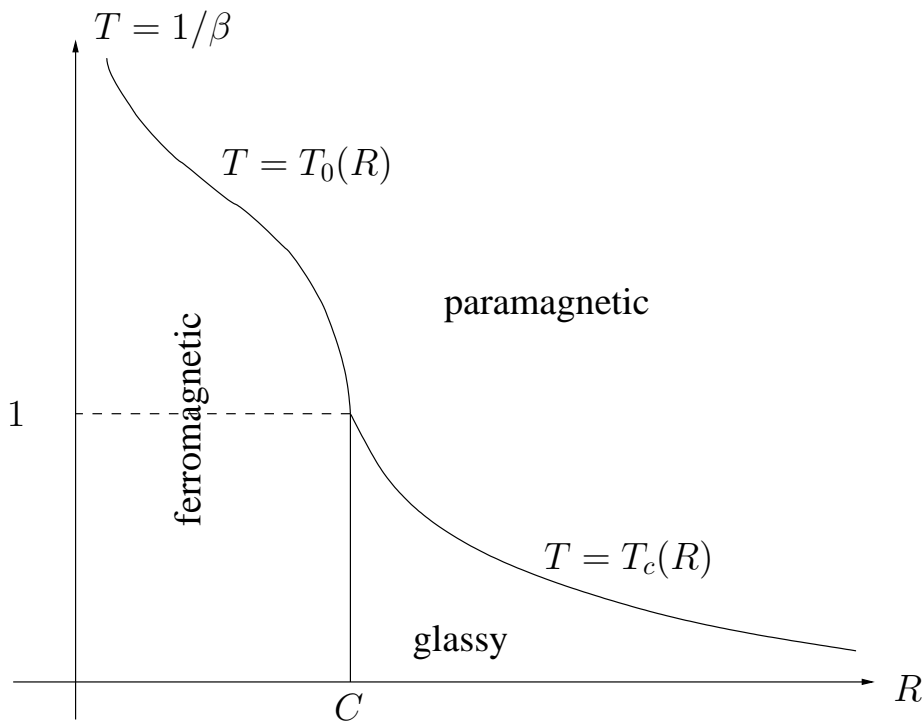}%
\end{picture}%
\setlength{\unitlength}{3947sp}%
\begingroup\makeatletter\ifx\SetFigFont\undefined%
\gdef\SetFigFont#1#2#3#4#5{%
  \reset@font\fontsize{#1}{#2pt}%
  \fontfamily{#3}\fontseries{#4}\fontshape{#5}%
  \selectfont}%
\fi\endgroup%
\begin{picture}(5569,4482)(751,-4244)
\put(2555,-4232){\makebox(0,0)[lb]{\smash{\SetFigFont{14}{16.8}{\rmdefault}{\mddefault}{\itdefault}{$~~~C$}%
}}}
\put(6320,-4114){\makebox(0,0)[lb]{\smash{\SetFigFont{14}{16.8}{\rmdefault}{\mddefault}{\itdefault}{$R$}%
}}}
\put(1300, 43){\makebox(0,0)[lb]{\smash{\SetFigFont{14}{16.8}{\rmdefault}{\mddefault}{\itdefault}{$T=1/\beta$}%
}}}
\put(751,-2388){\makebox(0,0)[lb]{\smash{\SetFigFont{14}{16.8}{\rmdefault}{\mddefault}{\itdefault}{$1$}%
}}}
\put(4126,-3286){\makebox(0,0)[lb]{\smash{\SetFigFont{14}{16.8}{\rmdefault}{\mddefault}{\itdefault}{$T=T_c(R)$}%
}}}
\put(2101,-811){\makebox(0,0)[lb]{\smash{\SetFigFont{14}{16.8}{\rmdefault}{\mddefault}{\itdefault}{$T=T_0(R)$}%
}}}
\end{picture}
\caption{Phase diagram of the finite--temperature MAP decoder.}
\label{gen1}
\end{figure}

In spite of the fact that in the glassy phase there are only few configurations
that dominate the behavior, it 
is no different from the paramagnetic phase in terms
of the typical ranking of the likelihood of the
correct codeword among all codewords: 
In both phases, the typical location of the
correct codeword in the list 
of descending likelihoods, $\{P(\by|\bx_i)\}$, is about
$2^{n(R-C)}$ ($R > C$). Although the glassy phase exhibits less uncertainty,
or equivalently, more certainty,
(sublinear conditional entropy given $\by$ about the channel
input), this relative certainty is misleading because the posterior probability
mass is captured mostly
by incorrect codewords. 
In this sense, the glassy phase is even more problematic
than the paramagnetic one: Since the certainty is fictitious,
it is more difficult to detect errors.

\subsubsection{Extension to General DMC's}

The extension to general DMC's is essentially quite straightforward.
Consider a DMC parametrized by $\{P(y|x),~x\in\calX,~
y\in\calY\}$. For the sake of simplicity, 
let us consider the uniform random coding
distribution\footnote{Other random coding distributions can be used
as well, but will lead to somewhat more complicated expressions, 
which we prefer to avoid in this description.}
according to which 
each codeword is selected independently at random with
probability distribution $Q(\bx)=1/|\calX|^n$ 
for all $\bx\in\calX^n$. For a given channel output
vector $\by$, the probability of 
selecting a random codeword $\bx$ whose conditional 
empirical distribution with $\by$ is $\hat{P}_{\bx|\by}$ 
is of the exponential order of $e^{-n[\ln|\calX|-\hat{H}_{\bx\by}(X|Y)]}$,
\cite{CK81}, thus the
expected number of codewords 
with this conditional distribution is exponentially
$$\bE\{N_{\by}(\hat{P}_{\bx|\by})\}\exe 
e^{n[R-\ln|\calX|+\hat{H}_{\bx\by}(X|Y)]}.$$ 
In analogy to the 
explanation provided in 
the previous subsection (and in \cite{MM06}), in the context of the BSC,
those conditional distributions $\{\hat{P}_{\bx|\by}\}$ for which
the exponent on the right--hand side is negative, are typically
not populated. Thus, for a typical random code
\begin{eqnarray}
Z_e(\beta|\by)&=&\sum_{\bx\in\calC-\{\bx_0\}}P^\beta(\by|\bx)\nonumber\\
&=&\sum_{\bx\in\calC-\{\bx_0\}}e^{-\beta\ln 1/P(\by|\bx)}\nonumber\\
&=&\sum_{\{\hat{P}_{\bx|\by}\}}N_{\by}(\hat{P}_{\bx|\by})
\cdot \exp\{-\beta\hat{\bE}_{\bx\by}\ln[1/P(Y|X)]\}\nonumber\\
&\exe&\exp\left\{n\left(R-\ln|\calX|+
\max_{Q_{X|Y}:~H_Q(X|Y)\ge \ln|\calX|-R}
\left[H_Q(X|Y)-\beta\bE_Q\{\ln[1/P(Y|X]\}\right]\right)\right\}\nonumber\\
&\dfn& e^{-n\beta F_e(\beta;Y)},
\end{eqnarray}
where $Y$ designates a RV distributed 
according to the empirical distribution $\hat{P}_{\by}$ of $\by$.

A word on notation is now in order: here and throughtout the sequel, 
we adopt the common abuse of notation, customarily used in the Information
Theory literature, that when a 
RV appears as an argument or a subscript of a certain 
function, this means that it is actually a functional of its distribution,
not a function of the value of the 
random variable itself. Whenever we wish to emphasize the
dependence of this quantity on the empirical distribution 
$\hat{P}_{\by}$, we will replace
$Y$ by $\hat{P}_{\by}$ or simply by $\by$ itself, provided that the context
does not leave room for ambiguity. Similar comments will apply to other
quantities to be defined throughout this 
subsection and in the sequel.\footnote{In Subsection 2.2.1, this 
issue did not arise since all relevant quantities happened to be independent
of $\hat{P}_{\by}$, due to the symmetry of the BSC.} For some of these
quantites, we will not 
denote the dependence on the distribution of 
$Y$ explicitly, in order to avoid cumbersome
notation, but it will be made clear that they do depend on it in general.

Consider now the expression
$$J_Y(\beta,R)\dfn\max_{Q_{X|Y}: H_Q(X|Y)\ge 
\ln|\calX|-R}\left[H_Q(X|Y)-\beta\bE_Q\{d(X,Y)\}\right],$$
where $d(x,y)\dfn -\ln p(y|x)$.

First, it is easy to prove (see Appendix, Subsection A.1)
that for fixed $\beta$ and $\by$, the function $J_Y(\beta,R)$
is concave in $R$. 
This means that the inequality 
constraint $H_Q(X|Y)\ge \ln|\calX|-R$ is met with equality
as long as $R \le R_Y(\beta)$,
where $R_Y(\beta)=\ln|\calX|-H_{Q_\beta}(X|Y)$ 
with $Q_\beta$ being the achiever of 
$$J_Y(\beta,\ln|\calX|)=\max_{Q_{X|Y}}[H_Q(X|Y)-\beta\bE_Q\{d(X,Y)\}],$$ 
that is,
$$Q_\beta(x|y)=\frac{e^{-\beta d(x,y)}}
{\sum_{x'\in\calX} e^{-\beta d(x',y)}}=\frac{P^\beta(y|x)}{\sum_{x'\in\calX}
P^\beta(y|x')}.$$
We will also use the notation 
$$D_Y(\beta)=\bE_{Q_\beta}\{d(X,Y)\}$$ 
and 
$$H_Y(\beta)=
H_{Q_\beta}(X|Y),$$ 
thus $R_Y(\beta)=\ln|X|-H_Y(\beta)$. 
Let
$$\beta_c(R)\dfn\inf\{\beta:~R_Y(\beta)\ge R\}=\inf\{\beta:~H_Y(\beta)\le \ln|\calX|-R\}.$$  
Obviously, $\beta_c(R)$ increases with $R$, or
equivalently, $T_c(R)=1/\beta_c(R)$ is decreasing with $R$
($T_c(\ln|\calX|)=0$). This forms the boundary
curve between the glassy and the paramagnetic phases. 
Note that when $R=I(X;Y)$, the mutual information
induced by the uniform distribution on $\calX$ and by $P(y|x)$, then
$\beta_c(R)=1$. Thus, $(I(X;Y),1)$ is a point on the curve $T=T_c(R)$.

For $R \le R_Y(\beta)$, or equivalently, $\beta\ge \beta_c(R)$,
the constraint $H_Q(X;Y)\ge \ln|\calX|-R$ is attained with equality.
Thus, in this range of low rates,
\begin{eqnarray}
J_Y(\beta,R)&=&\max_{\{Q_{X|Y}: H_Q(X|Y)
=\ln|\calX|-R\}}[\ln|\calX|-R-\beta\bE_Q\{d(X,Y)\}]\nonumber\\
&=&\ln|\calX|-R-\beta\cdot 
\min_{\{Q_{X|Y}: H_Q(X|Y)=\ln|\calX|-R\}}\bE_Q\{d(X,Y)\}\nonumber\\
&=&\ln|\calX|-R-\beta D_Y(\beta_R)
\end{eqnarray}
where $\beta_R$ is the solution to the equation $H_Y(\beta)=\ln|\calX|-R$.
We will also use the notation $\delta_Y(R)=D_Y(\beta_R)$.\footnote{The quantity 
$\delta_Y(R)$ 
is the generalization of the GV distance that
was defined in Subsection 2.2.1.\ for the BSC.}
It follows then
that $F_e(\beta,Y)=F_g(Y)=\delta_Y(R)$, which is the glassy phase.

For $R > R_Y(\beta)$, 
$$J_Y(\beta,R)=J_Y(\beta,\ln|\calX|)=\max_{Q_{X|Y}}
[H_Q(X|Y)-\beta\bE_Q\{d(X,Y)\}]=H_Y(\beta)-\beta D_Y(\beta)$$
Thus, for $\beta < \beta_c(R)$,
$$F_e(\beta,Y)=F_p(\beta,Y)=D_Y(\beta)+\frac{\ln|\calX|-R-H_Y(\beta)}{\beta},$$
which is the paramagnetic phase. It should be pointed out that for a general decoding metric
$d(x,y)$ (not necessarily ML matched to the channel), the boundary
between the paramagnetic and the 
glassy phases depends only on the random coding distribution
and this decoding metric 
$d(x,y)$, not on the channel itself (cf.\ Subsection 2.2.3). 
The boundaries with the
ferromagnetic phase are the ones that depend on the channel.

In the ordered (ferromagnetic) phase, the free energy density is given by
$F(\beta)=H(Y|X)$, where $X$ is uniform and $Y$ given $X$ is 
distributed according to
the channel. As long as $R < I(X;Y)$, we have $H(Y|X) < \delta_Y(R)$. 
In fact, the line connecting the points
$(R=I(X;Y),T=1)$ and $(R=I(X;Y),T=0)$ forms the boundary between the
ordered ferromagnetic phase and the glassy phase.

For $R < I(X;Y)$, the boundary 
between the ferromagnetic and paramagnetic phases is given by
the solution $\beta_0(R)$ (or $T_0(R)=1/\beta_0(R)$) to the equation
$$\beta H(Y|X)=\beta D_Y(\beta)+\ln|\calX|-R-H_Y(\beta),$$
which is above the curve $T=T_c(R)$ for $R < I(X;Y)$. 
It should be emphasized that
$\beta_c(R)$, $\beta_0(R)$, 
and $\beta_R$ all depend on the (distribution of the) RV $Y$,
namely, the empirical distribution of $\by$.

\subsubsection{Phase Diagram for Universal Decoding}

It is instructive to compare 
the phase diagram of finite--temperature MAP decoding to 
those of finite--temperature universal decoders.
One simple example of a universal decoder for which it is especially
easy to derive the phase diagram is the minimum conditional entropy
decoder \cite{Ziv85}, which given $\by$, selects the codeword $\bx_m$ for which
$\hat{H}_{\bx_m\by}(X|Y)$ is minimum.\footnote{This is a variant of the
well--known maximum mutual information (MMI) decoder.
In the case of constant composition codes, these two decoders
are identical.} It is well known
that this universal decoder is asymptotically optimum in the random coding
sense, in that it achieves the same random coding error exponent as the
ML decoder, provided that the random coding 
distribution is uniform over $\calX^n$.

The partition function corresponding to this universal decoder is the
same as before, except that
$\hat{\bE}_{\bx\by}\{d(X,Y)\}$ 
is replaced by $\hat{H}_{\bx\by}(X|Y)$.
In this case,
\begin{eqnarray}
Z_e(\beta|\by)
&=&\sum_{\{\hat{P}_{\bx|\by}\}}
N_{\by}(\hat{P}_{\bx|\by})\cdot e^{-\beta \hat{H}_{\bx\by}(X|Y)}\nonumber\\
&\exe&\exp\left\{n\left(R-\ln|\calX|+
\max_{Q_{X|Y}:~H_Q(X|Y)\ge \ln|\calX|-R}
[(1-\beta)H_Q(X|Y)]\right)\right\}\nonumber\\
&\dfn& e^{-n\beta F_e(\beta,Y)}
\end{eqnarray}
Now, it is easy to see how 
phase transitions behave (see Fig.\ 2): If $\beta < 1$, then
the maximum is $\ln|\calX|$ and we get 
$$Z_e(\beta)\exe e^{n[R-\beta\ln |\calX|]},$$
thus $F_e(\beta,Y)=F_p(\beta)=\ln|\calX|-R/\beta$. If $\beta > 1$, we get
$$Z_e(\beta)\exe e^{-n\beta[\ln |\calX|-R]},$$
thus, $F_e(\beta,Y)=F_g=\ln|\calX|-R$. 
Therefore, the boundary between the two phases is the
horizontal line $T_c=1/\beta_c=1$ 
(independently of $R$). This means that the glassy
region here is larger than in ML decoding for $R > C$.
The boundary between the ferromagnetic
and the glassy phases continues to be $R=I(X;Y)$ 
as before. The ferromagnetic--paramagnetic boundary is now
$H(X|Y)=\ln|\calX|-R/\beta$, or, equivalently, 
$T=1/\beta= I(X;Y)/R$, which is below the
ferromagnetic--paramagnetic
boundary of the MAP decoder.
This can easily be shown by setting $R=\beta I(X;Y)$
(which is this boundary) in the 
r.h.s.\ of the equation defining $T_0(R)$ and showing that
the resulting expression is larger than $\beta H(Y|X)$ (for $\beta \le 1$), 
which is the l.h.s.\ of this equation
(thus, we are still in the ferromagnetic phase 
of MAP decoding): Specifically,
the l.h.s.\ of the equation defining $T_0(R)$ is:
\begin{eqnarray}
& &\beta D_Y(\beta)+\ln|\calX|-R-H_Y(\beta)\nonumber\\
&=&\beta D_Y(\beta)+\ln|\calX|-\beta I(X;Y)-H_Y(\beta)\nonumber\\
&=&\beta H(Y|X)+\beta \bE_{Q_\beta}\ln\frac{1}{P(Y|X)}+
\ln|\calX|-\beta H(Y)-H_{Q_\beta}(X|Y)\nonumber\\
&=&\beta H(Y|X)+\beta \bE_{Q_\beta}\ln\frac{1}{P(Y|X)}
-\beta H(Y)+I_{Q_\beta}(X;Y)\nonumber\\
&\ge&\beta H(Y|X)+\beta H_{Q_\beta}(Y|X)
-\beta H(Y)+I_{Q_\beta}(X;Y)\nonumber\\
&\ge&\beta H(Y|X)-\beta I_{Q_\beta}(X;Y)
+I_{Q_\beta}(X;Y)\nonumber\\
&\ge& \beta H(Y|X)
\end{eqnarray}
where the first equality is since 
$R=\beta I(X;Y)$ on the boundary, and the last equality is since $\beta\le 1$.
Thus, although this decoder achieves the optimum random coding 
error exponent, it has a phase diagram which is worse than that
of MAP decoding, as the ferromagnetic 
region is smaller and the glassy region is larger.

\begin{figure}[ht]
\hspace*{1cm}
\begin{picture}(0,0)%
\includegraphics{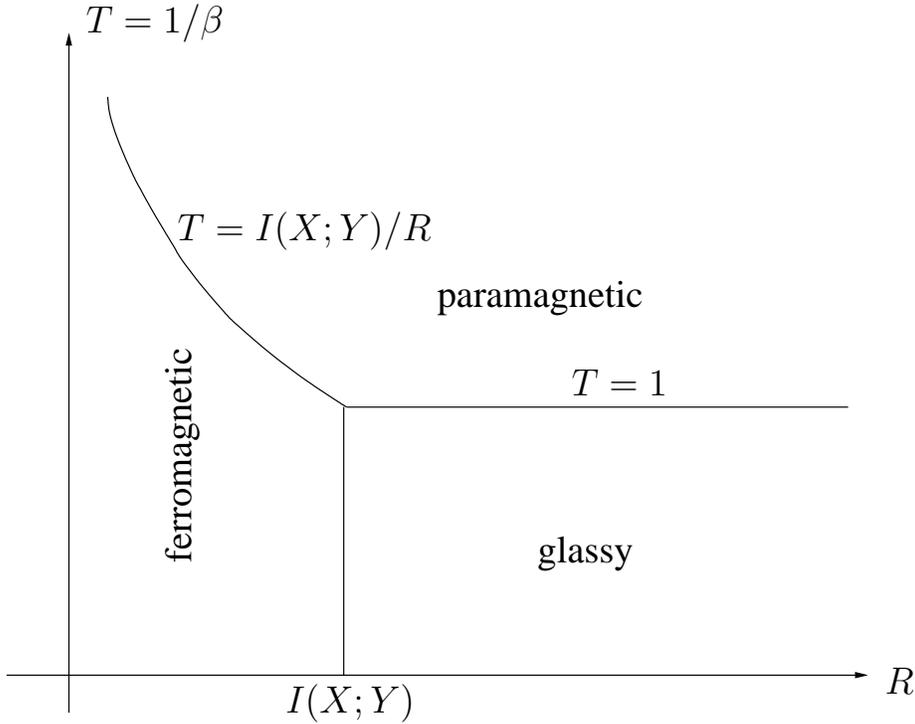}%
\end{picture}%
\setlength{\unitlength}{3947sp}%
\begingroup\makeatletter\ifx\SetFigFont\undefined%
\gdef\SetFigFont#1#2#3#4#5{%
  \reset@font\fontsize{#1}{#2pt}%
  \fontfamily{#3}\fontseries{#4}\fontshape{#5}%
  \selectfont}%
\fi\endgroup%
\begin{picture}(5542,4482)(778,-4244)
\put(2555,-4232){\makebox(0,0)[lb]{\smash{\SetFigFont{14}{16.8}{\rmdefault}{\mddefault}{\itdefault}{$I(X;Y)$}%
}}}
\put(6320,-4114){\makebox(0,0)[lb]{\smash{\SetFigFont{14}{16.8}{\rmdefault}{\mddefault}{\itdefault}{$R$}%
}}}
\put(1300, 43){\makebox(0,0)[lb]{\smash{\SetFigFont{14}{16.8}{\rmdefault}{\mddefault}{\itdefault}{$T=1/\beta$}%
}}}
\put(4351,-2236){\makebox(0,0)[lb]{\smash{\SetFigFont{14}{16.8}{\rmdefault}{\mddefault}{\itdefault}{$T=1$}%
}}}
\put(1876,-1261){\makebox(0,0)[lb]{\smash{\SetFigFont{14}{16.8}{\rmdefault}{\mddefault}{\itdefault}{$T=I(X;Y)/R$}%
}}}
\end{picture}
\caption{Phase diagram for universal decoding.}
\label{gen2}
\end{figure}

\section{The Correct Decoding Exponent}

We now proceed to establish 
relationships between the phase diagram of a random code,
decoded by a finite temperature MAP decoder,
and the exponent of correct decoding at rates above capacity, or to 
be more precise, rates above $I(X;Y)$, the mutual information
induced by the uniform input distribution and the channel.

Arimoto \cite{Arimoto73} begins the derivation of his bound on 
the probability of correct decoding by using the inequality
\begin{equation}
P_c=\frac{1}{M}\sum_{\by\in\calY^n}\max_{1\le i\le M} P(\by|\bx_i)
\le\frac{1}{M}\sum_{\by\in\calY^n}
\left[\sum_{i=1}^M P^\beta(\by|\bx_i)\right]^{1/\beta},~~~~\beta > 0
\end{equation}
which becomes tight when $\beta\to\infty$. We will also use this inequality,
but we shall proceed somewhat differently than Arimoto. First, observe that
for a randomly selected code, where
the average probability of correct decoding is upper bounded by
\begin{equation}
\bar{P}_c
\le\frac{1}{M}\sum_{\by\in\calY^n}\bE\left\{
\left[\sum_{i=1}^M P^\beta(\by|\bx_i)\right]^{1/\beta}\right\},
\end{equation}
the expression in the square brackets is exactly $Z_e(\beta)$
(just with $M$ codewords instead of $M-1$), because the
interpretation of this expression, is that
the codewords are drawn under $Q$ regardless of $\by$.
Since we are interested in
$\beta\to \infty$ (in addition to the assumption that $R>I(X;Y)$),
then we are actually carrying out this calculation in the {\it glassy} regime.

The above upper bound to $\bar{P}_c$ can be also written as:
\begin{equation}
\bar{P}_c\le 
\frac{1}{M}\sum_{\by\in\calY^n}
\bE\left\{\left[\sum_{d\in \calD_n} 
N_{\by}(d)\cdot e^{-\beta d}\right]^{1/\beta}\right\},
\end{equation}
where here $N_{\by}(d)$ denotes the number of codewords $\bx_i$ for which
$-\ln P(\by|\bx_i)=d$, and $\calD_n$ is the set
of values that the function
$d(\bx,\by)=-\ln P(\by|\bx)$ 
can take on for a given $\by$,
as $\bx$ exhausts the codebook $\calC$.
Note that as $d(\bx,\by)$ depends
only on the empirical 
joint distribution of $\bx$ and $\by$, then $|\calD_n|$ cannot
exceed the number of empirical 
conditional distributions (or conditional type classes) 
corresponding to pairs of 
$n$--sequences, and so, $|\calD_n|$ is upper bounded by a polynomial in $n$.

Now, when a random code is considered, 
then instead of applying Jensen's inequality 
for $\beta\ge 1$ (as was done in \cite{Arimoto73}),
and thereby insert the expectation 
operator into the square brackets, let us adopt another approach.
Consider the following events:
$$\calB=\left\{ \calC:~N_{\by}(d)\ge
\exp\{n[R-\ln|\calX|+h_0(d/n|\by)]_++\epsilon]\}~
\mbox{for some $d\in\calD_n$}\right\},$$
where $[t]_+\dfn=\max\{0,t\}$ and
where $h_0(\delta|\by)$ is defined as the
maximum of $H_Q(X|Y)$ subject to the constraints that $\bE_Q\{d(X,Y)\}=\delta$
and that $Y$ is distributed according to $\hat{P}_{\by}$.
Also, define
$$\calW_i =\left\{ \calC:~\min\{d:~N_{\by}(d)\ge
1\}=i\right\},~~~i\le d_0(\by)\dfn n\delta_Y(R),$$
where we recall that
$\delta_Y(R)$ is the solution to the equation $h_0(\delta|\by)=\ln|\calX|-R$.
Note that $\{\calW_i\}$ are disjoint events.
Now, for $\beta > \beta_c(R)$:
\begin{eqnarray}
&& \bE\left\{\left[\sum_{d\in\calD_n} N_{\by}(d)\cdot
e^{-\beta d}\right]^{1/\beta}\right\}\nonumber\\
&\le&\mbox{Pr}\{\calB\}\cdot[e^{nR}\cdot
e^{-\beta \cdot 0}]^{1/\beta}+\nonumber\\
& &+\sum_{d\le d_0(\by)}\mbox{Pr}\{\calW_d\cap\calB^c\}\cdot
\left[e^{n\epsilon}e^{-\beta d}\right]^{1/\beta}+\nonumber\\
& &+\mbox{Pr}\{\calW_0^c\cap\calW_1^c\cap\ldots
\cap\calW_{d_0(\by)}^c\cap\calB^c\}\cdot
e^{-nF_g(Y)}\cdot e^{n\epsilon/\beta},
\end{eqnarray}
This inequality calls for some explanation: 
We are dividing the set of configurations of the RV's $\{N_{\by}(d)\}_{d\in
\calD_n}$ into three classes, defined by the events
$\calB$ and $\{\calW_i\}$. In the first class, corresponding to the
first term on the right--hand side,  $\{N_{\by}(d)\}$ fall in
$\calB$, where there is at least one value of $d$ for which
$N_{\by}(d)$ is {\it exponentially} larger (by at least $\epsilon$) 
than its expectation.
We bound the value of $[\sum_{d\in\calD_n} N_{\by}(d)\cdot 
e^{-\beta d}]^{1/\beta}$, 
in this class, very ``generously'', by the maximum
possible value it can possibly
take, that is, when all $e^{nR}$ codewords are at zero
distance from $\by$, but this quantity is weighted by $\mbox{Pr}\{\calB\}$,
which as is shown in the Appendix (Subsection A.2),
decays double--exponentially
rapidly, at least as fast as $e^{-e^{n\epsilon}}$,
and so this first term is negligible. The other two classes correspond
to $\calB^c$, where for all $d\in\calD_n$, $N_{\by}(d)$ does not exceed
its expectation times $e^{n\epsilon}$. Here we distinguish between two
cases (corresponding to the two other classes): In one of them,
(at least) one of the distances below the generalized GV distance
$d_0(\by)=n\delta_Y(R)$ is populated by subexponentially\footnote{The event
$\calB^c$ guarantees that there are only subexponentially many codewords
at distances below $d_0(\by)$.} many 
codewords. Since we are operating in the glassy
regime, the dominant contribution to 
$[\sum_{d\in\calD_n} N_{\by}(d)\cdot 
e^{-\beta d}]^{1/\beta}$ will be due to
these minimum distance codewords, and the weighting 
of the event of minimum distance $d$ is, of course,
according to $\mbox{Pr}\{\calW_d\cap\calB^c\}$. In the other case,
which captures most of the probability mass (since it is the typical
configuration of $\{N_{\by}(d)\}$),
none of the distances below the generalized GV distance is populated
by codewords, whereas for larger distances, $\{N_{\by}(d)\}$ are all
(within a factor of $e^{n\epsilon}$) about their expectations. In this
case, our expression again behaves according to the glassy regime, where
the generalized GV distance dominates the partition function.

Now, regarding the second term, for $\delta =d/n < \delta_Y(R)$,
\begin{equation}
\mbox{Pr}\{\calW_d\cap\calB^c\}\le
\mbox{Pr}\{\calW_d\}\le \mbox{Pr}\{N_{\by}(d)\ge 1\},
\end{equation}
where the latter expression is shown 
(Appendix, Subsection A.2) to decay at the exponential rate of
$e^{-n[\ln|\calX|-R-h_0(\delta|\by)]}$.
Thus,
\begin{eqnarray}
&& \bE\left\{\left[\sum_{d\in\calD_n} N_{\by}(d)\cdot
e^{-\beta d}\right]^{1/\beta}\right\}\nonumber\\
&\le& e^{-e^{n\epsilon}}\cdot[e^{nR}]^{1/\beta}+\nonumber\\
& &+\sum_{\delta\le \delta_Y(R)} e^{-n[\ln|\calX|-R-h_0(\delta|\by)]}\cdot
\left[e^{n\epsilon}e^{-\beta n\delta}\right]^{1/\beta}+
e^{-nF_g(Y)}\cdot e^{n\epsilon/\beta}\nonumber\\
&\exe& e^{n(R-\ln|\calX|)}\cdot e^{n\epsilon/\beta}
\exp\{n\max_{\delta\le\delta_Y(R)}
[h_0(\delta|\by)-\delta]\}
+e^{-nF_g(Y)}\cdot e^{n\epsilon/\beta}\nonumber\\
&\exe& e^{n(R-\ln|\calX|)}\cdot e^{n\epsilon/\beta}
\exp\{n[h_0(\delta_Y(R)|\by)-\delta_Y(R)]\}
+e^{-nF_g(Y)}\cdot e^{n\epsilon/\beta}\nonumber\\
&\exe& e^{n(R-\ln|\calX|)}\exp\{n[\ln|\calX|-R-\delta_Y(R)]\}
\cdot e^{n\epsilon/\beta}
+e^{-nF_g(Y)}\cdot e^{n\epsilon/\beta}\nonumber\\
&\exe&e^{-nF_g(Y)}\cdot e^{n\epsilon/\beta}.
\end{eqnarray}
Since $\epsilon$ can be chosen arbitrarily small for large $n$
(in fact, one may let $\epsilon$ vanish with $n$ sufficiently slowly),
the exponential rate of the expression under discussion
is actually bounded by $e^{-nF_g(Y)}$.
Note that whenever $\beta \ge \beta_c$,
this expression no longer depends on $\beta$. Finally,
substituting this bound back into the bound on $\bar{P}_c$, we get:
\begin{eqnarray}
\bar{P}_c&\le&\frac{1}{M}\sum_{\by\in\calY^n} e^{-nF_g(Y)}\nonumber\\
&\exe&e^{-nR}\cdot e^{n\max_Y[H(Y)-F_g(Y)]}\nonumber\\
&=&e^{-n(R-\max_Y[H(Y)-F_g(Y)])}.
\end{eqnarray}
This calculation can be shown to be
exponentially tight: a lower bound can be obtained by confining
the calculation to the (high probability) event
$\calW_0^c\cap\calW_1^c\cap\ldots\cap\calW_{d_0(\by)}^c\cap\calB^c$
with the additional restriction that 
$N_{\by}(d)\ge \bE\{N_{\by}(d)\}\cdot e^{-n
\epsilon}$
for all $d\ge d_0(\by)$
(i.e., the last term only in the above derivation). Note that in
Arimoto's paper, where Jensen's inequality is used, the expectation
of $\sum_d N_{\by}(d)e^{-\beta d}$ is computed,
and this actually
corresponds to the
paramagnetic regime (without the constraint $H_Q(X|Y)\ge \ln|\calX|-R$).
The resulting bound 
might not be exponentially tight in general.\footnote{Note that although
the exact reliability function 
(for optimum codes) for rates above capacity was established
by Dueck and K\"orner \cite{DK79}, 
here we are only focusing on random codes drawn under an
i.i.d.\ distribution.}
Finally, the optimization
$\max_Y[H(Y)-F_g(Y)]$ can be carried out explicitly, yielding
$\ln\sum_y e^{-f_g(y)}$, where 
where $f_g(y)=\bE_{Q_{\beta_R}}\{d(X,Y)|Y=y\}$.

We have obtained then a random coding exponent formula in terms
of the free energy density in the glassy phase, from which we learn that
the free energy density of the glassy phase plays a central role in
the calculation the exponent of correct decoding. To obtain some insight,
it is instructive to examine this expression in the special case of the BSC.
Here, since 
$F_g=F_g(Y)$ does not depend on the probability distribution of $Y$,
we get:
\begin{eqnarray}
\bar{P}_c &\le&e^{n[\ln 2-R-F_g]}\nonumber\\
&=&e^{n[h(\delta_{GV}(R))-\delta_{GV}(R)\ln\frac{1}{p}-
(1-\delta_{GV}(R))\ln\frac{1}{1-p}]}\nonumber\\
&=&e^{-nD(\delta_{GV}(R)\|p)},
\end{eqnarray}
where for $a,b\in(0,1)$, 
$D(a\|b)\dfn a\ln\frac{a}{b}+(1-a)\ln\frac{1-a}{1-b}$.
This result has the intuitively appealing interpretation of 
the probability of the large deviations event 
that the channel makes $n\delta_{GV}(R)$
errors or less, although $p >\delta_{GV}(R)$), 
in which case the correct codeword `penetrates'
into the sphere of radius $n\delta_{GV}(R)$,
whose surface is populated by the codewords that dominate
the glassy phase. Of course, when such an event happens, the correct
codeword dominates the partition function, and thus the
decoding is correct.

\section{The Random Coding Error Exponent}

Let us now examine rates below $I(X;Y)$.
Consider Gallager's upper bound on the error probability
for a given code \cite{VO79}:
\begin{equation}
P_e\le \frac{1}{M}\sum_{m=1}^M\sum_{\by\in\calY^n}P(\by|\bx_m)^{1/(1+\rho)}
\cdot\left[\sum_{m'\ne m} 
P(\by|\bx_{m'})^{1/(1+\rho)}\right]^\rho ~~~~\rho\ge 0.
\end{equation}
The bracketed term is once 
again identified with $Z_e(\beta)$ for $\beta=\frac{1}{1+\rho} \le 1$,
in contrast to the calculation of $P_c$, where we used large
values of $\beta$. For each $m$, 
let us first take only the expectation w.r.t.\ the incorrect
codewords, referring to the random variables $\{N_{\by}(d)\}$. 
Let this partial expectation be denoted by $\tilde{P}_e$. 
We will also denote $\frac{1}{1+\rho}$ by $\beta$. 
One way to carry out this calculation is to use the same technique
as we used in the previous section, by classifying the distance
spectrum $\{N_{\by}(d)\}$ to its various classes. However, here
since we know already that the use of Jensen's inequality would not harm
the exponential tightness \cite{Gallager73}, it will be simpler to
apply Jensen's inequality (for $0\le\rho\le 1$, that is,
$0.5\le\beta\le 1$) 
and thereby essentially carry out the calculation 
in the {\it paramagnetic} regime.
We proceed then as follows:
\begin{eqnarray}
\tilde{P}_e&\le&\frac{1}{M}\sum_{m=1}^M
\sum_{\by\in\calY^n}P(\by|\bx_m)^\beta\cdot
\bE\left\{\left[\sum_{d\in\calD_n} 
N_{\by}(d)e^{-\beta d}\right]^\rho\right\}\nonumber\\
&\le&\frac{1}{M}\sum_{m=1}^M\sum_{\by\in\calY^n}P(\by|\bx_m)^\beta\cdot
\left[\sum_{d\in\calD_n} \bE\{N_{\by}(d)\}
\cdot e^{-\beta d}\right]^\rho\nonumber\\
&\exe&\frac{1}{M}\sum_{m=1}^M\sum_{\by\in\calY^n}P(\by|\bx_m)^\beta\cdot
\left[\sum_{d\in\calD_n} e^{n[R-\ln|\calX| 
+h_0(\delta|\by)]}\cdot e^{-\beta d}\right]^\rho\nonumber\\
&\exe&\frac{1}{M}\sum_{m=1}^M\sum_{\by\in\calY^n}P(\by|\bx_m)^\beta\cdot
[e^{-n\beta F_p(\beta,Y)}]^\rho.
\end{eqnarray}
Next, we take the expectation w.r.t.\ the correct codeword $\bx_m$: Define
$$\Gamma(y)=\ln\sum_{x\in\calX}P^\beta(y|x)-\ln|\calX|,~~~y\in\calY.$$
Then, the average error probability $\bar{P}_e$ is upper bounded by
\begin{eqnarray}
\bar{P}_e&\le&\sum_{\by\in\calY^n} e^{\sum_{i=1}^n \Gamma(y_i)}
\cdot e^{-n\rho\beta F_p(\beta,Y)}\nonumber\\
&=&\sum_{\by\in\calY^n} \exp\{n[\hat{\bE}_{\by}
\Gamma(y)-\rho\beta F_p(\beta,Y)]\}\nonumber\\
&\exe&\exp\{n\cdot\max_Y[H(Y)+
\sum_{y\in\calY}P(y)\Gamma(y)-\rho\beta F_p(\beta,Y)]\}\nonumber\\
&=&\exp\{-n\cdot\min_Y[\rho\beta F_p(\beta,Y)-
\sum_{y\in\calY}P(y)\Gamma(y)-H(Y)]\}.
\end{eqnarray}
Note that $\Gamma(y)$ is 
also related to a free energy expression, corresponding
to the uniform
prior over the entire input space $\calX^n$, not only the codebook.
Thus, we have two free energy expressions, one pertaining to the
contribution of the correct codeword, and the other is related to
the contributions of the incorrect codewords.

In the special case of the BSC, 
where $F_p(\beta,Y)$ and does not depend on $Y$ and
$\Gamma=\Gamma(y)$ does not depend on $y$, we get
the exponential rate of 
\begin{eqnarray}
&&\min_Y[\beta\rho F_p(\beta)-\Gamma-H(Y)]\nonumber\\
&=&\beta\rho F_p(\beta)-(\ln[p^\beta+(1-p)^\beta]-\ln 2)-\ln 2\nonumber\\
&=&\rho(\ln 2-R)-(1+\rho)\ln[p^\beta+(1-p)^\beta]\nonumber\\
&=&\rho \ln 2-(1+\rho)\ln[p^{1/(1+\rho)}
+(1-p)^{1/(1+\rho)}]-\rho R\nonumber\\
&=& E_0(\rho)-\rho R\nonumber\\
&\dfn& E_0(\rho,R)
\end{eqnarray}
which is, as expected, Gallager's reliability function for the BSC. 
The optimum choice of $\rho$ depends on $R$. 
As is shown in \cite[pp.\ 151-152]{VO79},
in the range $R \le\ln 2- h(p_{1/2})$, 
that is, $p_{1/2} < \delta_{GV}(R)$, we have
$\rho=1$, which means $\beta=\frac{1}{2}$. 
For $R\in [\ln 2-h(p_{1/2}),\ln 2-h(p)]$, the 
optimum $\rho$ is in $[0,1)$, and it satisfies
$R=\ln 2-h(p_{1/(1+\rho)})=\ln 2-h(p_\beta)$, 
or, equivalently, $p_\beta=\delta_{GV}(R)$,
which means that we move along the boundary
between the the glassy phase and the paramagnetic phases of $Z_e(\beta|\by)$.

\clearpage
\section*{Appendix}

\subsection*{A.1 Proof of the Concavity of $J_Y(\beta,\cdot)$}

Let $Q_1$ and $Q_2$ achieve $J_Y(\beta,R_1)$ and $J_Y(\beta,R_2)$,
respectively. Now, let $Q=\alpha Q_1+(1-\alpha)Q_2$ for some $\alpha\in(0,1)$.
First, observe that by the concavity of the conditional entropy in $Q_{X|Y}$
for fixed $Q_Y$, we have 
$$H_Q(X|Y)\ge\alpha H_{Q_1}(X|Y)+(1-\alpha)H_{Q_2}(X|Y)\ge \ln|\calX|-\alpha R_1-(1-\alpha)R_2.$$
It follows then that $H_Q(X|Y)-\beta\bE_Qd(X,Y)\le J(\beta,\alpha R_1+(1-\alpha)R_2|\by)$.
But, on the other hand
\begin{eqnarray}
H_Q(X|Y)-\beta\bE_Qd(X,Y)&\ge&\alpha[H_{Q_1}(X|Y)-\beta\bE_{Q_1}d(X,Y)]+(1-\alpha)
[H_{Q_2}(X|Y)-\beta\bE_{Q_2}d(X,Y)]\nonumber\\
&=&\alpha J_Y(\beta,R_1)+(1-\alpha)J_Y(\beta,R_2).
\end{eqnarray}
Thus,
$$J_Y(\beta,\alpha R_1+(1-\alpha)R_2)\ge \alpha J_Y(\beta,R_1)+(1-\alpha)J_Y(\beta,R_2).$$

\subsection*{A.2 Large Deviations Behavior of $N_{\by}(d)$}

For $a,b\in[0,1]$, consider the binary divergence
\begin{eqnarray}
D(a\|b)&\dfn&a\ln \frac{a}{b}+(1-a)\ln\frac{1-a}{1-b}\nonumber\\
&=&a\ln \frac{a}{b}+(1-a)\ln\left[1+\frac{b-a}{1-b}\right]
\end{eqnarray}
To derive a lower bound to $D(a\|b)$, let us use the inequality
$$\ln(1+x)=-\ln\frac{1}{1+x}=-\ln\left(1-\frac{x}{1+x}\right)\ge \frac{x}{1+x},$$
and then
\begin{eqnarray}
D(a\|b)&\ge&a\ln \frac{a}{b}+(1-a)\cdot\frac{(b-a)/(1-b)}
{1+(b-a)/(1-b)}\nonumber\\
&=&a\ln \frac{a}{b}+b-a\nonumber\\
&>&a\left(\ln\frac{a}{b}-1\right).
\end{eqnarray}
Now, let $N_{\by}(d)$ denote the number of codewords for
which $-\ln P(\by|\bx_i)=d$.
As mentioned earlier, $N_{\by}(d)$ is the sum of the $e^{nR}$ 
independent binary random variables $1\{d(\bX_i,\by)=d\}$,
where the probability that $d(\bX_i,\by)=d$ is 
exponentially $b=e^{-n[\ln|\calX|- h_0(\delta|\by)]}$, $h_0(\delta|\by)$ being
the maximum of $H_Q(X|Y)$ subject to the constraints that 
$\bE_Q\{d(X,Y)\}=\delta$, $\delta=d/n$, and that $Y$ is distributed according
to $\hat{P}_{\by}$.
The event $N_{\by}(d)\ge e^{nA}$, for $d=\delta n$ and $A\in[0,R)$,
means that the relative frequency of the event $1\{d(\bX_i,\by)=d\}$
is at least $a=e^{-n(R-A)}$. Thus, by the Chernoff bound:
\begin{eqnarray}
\mbox{Pr}\{N_{\by}(d)\ge e^{nA}\}&\le&
\exp\left\{-e^{nR}D(e^{-n(R-A)}\|e^{-n[\ln|\calX|-
h_0(\delta|\by)]})\right\}\nonumber\\
&\le&
\exp\left\{-e^{nR}\cdot e^{-n(R-A)}(n[(\ln|\calX|-R-
h_0(\delta|\by)+A]-1)\right\}\nonumber\\
&\le&
\exp\left\{-e^{nA}(n[\ln|\calX|-R-h_0(\delta|\by)+A]-1)\right\}.
\end{eqnarray}
Now, for $A=[R-\ln|\calX|+h_0(\delta|\by)]_++\epsilon$, 
the term in the square brackets is at least
$\epsilon > 0$, and thus $\mbox{Pr}\{N_{\by}(d)\ge e^{nA}\}$ 
decays double--exponentially rapidly,
not slower than $e^{-e^{n\epsilon}}$.
The probability of the union of the (polynomially many) events 
$\{N_{\by}(d)\ge e^{nA}\}_{d\in\calD_n}$, which is
upper bounded by the sum of the probabilities,
is still double--exponentially small.
Thus, $\mbox{Pr}\{\calB\}$ decays 
double--exponentially rapidly. Now, the event
$\{N_{\by}(d)\ge 1\}$ corresponds to the 
choice $A=0$. For $\delta < \delta_Y(R)$,
$\delta_Y(R)$ being the solution to the
equation $\ln|\calX|-R=h_0(\delta|\by)$, 
which means that $\ln|\calX|-R-h_0(\delta|\by) > 0$, 
this gives an ordinary exponential decay
at the rate of $e^{-n[\ln|\calX|-R-h_0(\delta|\by)]}$.

\end{document}